\documentclass[aps,10pt,prd,notitlepage,showpacs,nofootinbib,superscriptaddress]{revtex4-2}
\usepackage{graphicx}
\usepackage[utf8]{inputenc} 
\usepackage{amsmath}
\usepackage{amssymb}
\usepackage{siunitx}
\usepackage[section]{placeins}
\usepackage{float}
\usepackage{comment}
\usepackage{slashed}
\usepackage[normalem]{ulem}
\usepackage{braket}
\usepackage{comment}

\usepackage{caption}
\usepackage{subcaption}

\usepackage[usenames,dvipsnames]{color}
\usepackage[colorinlistoftodos]{todonotes}
\usepackage[colorlinks=true,citecolor=blue,urlcolor=magenta, pdfborder={0 0 0}]{hyperref}

\usepackage{soul}

\begin{document}
\bibliographystyle{unsrt}   

\title{Dark matter production from two evaporating PBH distributions}

\author{Arnab Chaudhuri}
\email{arnab.c@iitgn.ac.in }
\affiliation{Indian Institute of Technology Gandhinagar, Gujarat 382055, India}

\author{Baradhwaj Coleppa}
\email{baradhwaj@iitgn.ac.in }
\affiliation{Indian Institute of Technology Gandhinagar, Gujarat 382055, India}

\author{Kousik Loho}
\email{kousik.loho@iitgn.ac.in}
\affiliation{Indian Institute of Technology Gandhinagar, Gujarat 382055, India}

\date{\today}
\begin{abstract}
Particulate Dark Matter (DM), completely isolated from the Standard Model particle sector, can be produced in the early universe from Primordial Black Hole (PBH) evaporation. However, Big Bang Nucleosynthesis (BBN) observations put an upper bound on the initial mass of PBH requiring the PBH to evaporate completely before the advent of BBN. DM particles in the  mass range $\sim(1-10^9)$ GeV can not explain the observed relic abundance for an early matter dominated universe due to this BBN constraint. However, this assumes the presence of only one monochromatic PBH mass distribution in the early universe. In this work, we explore the simple possibility of achieving the observed relic with DM masses from the above mentioned range for an early matter dominated era with two monochromatic evaporating PBH mass distributions and demonstrate that the fermionic DM masses consistent with BBN change slightly.
\end{abstract}
\maketitle

\section{Introduction}
\label{sec:intro}

The Standard Model (SM) of particle physics continues to be the most successful theory that describes the physics of elementary particles and interactions barring the gravitational ones. However, there are certain issues with the SM, of which a very important one is that the SM can not explain the particulate nature of Dark Matter (DM). The presence of DM is evident in the universe \cite{Clowe:2006eq,Sofue:2000jx,Metcalf:2003sz,Bartelmann:2010fz} and it accounts for more than one fourth of the energy budget of the universe \cite{Planck:2018vyg, WMAP:2012nax}. Some of the popular mechanisms of DM production are the WIMP (weakly interacting massive particle) \cite{Bertone:2004pz,Bergstrom:2009ib,Arcadi:2017kky,Bauer:2017qwy} and FIMP (feebly interacting massive particle) \cite{Hall:2009bx,Bernal:2017kxu} scenarios which require the DM candidate(s) to have a portal coupling with the SM sector. However, a DM particle, completely isolated from the SM sector, can be produced from the Primordial Black Hole (PBH) evaporation via Hawking radiation \cite{Hawking:1974rv,Hawking:1975vcx} and is hence immune to various direct detection \cite{XENON:2018voc,LUX:2016ggv,PandaX-II:2017hlx} and collider constraints \cite{Boveia:2018yeb}. 

PBHs can be produced in the early radiation dominated universe from quantum fluctuations among other production mechanisms mentioned in section \ref{sec:pbh} and later on dominate the energy density of the universe only to finally evaporate away completely by radiating DM particles along with other SM states before the advent of the Big Bang Nucleosynthesis (BBN). The phenomenology of DM production from PBH evaporation in the context of a single PBH initial mass has been widely studied in the literature 
\cite{Baldes:2020nuv,Bernal:2020ili,Dai:2009hx,Fujita:2014hha,Morrison:2018xla,Hooper:2019gtx,Bernal:2020kse,JyotiDas:2021shi,Gondolo:2020uqv,Masina:2020xhk,Borah:2022vsu,Lennon:2017tqq,Coleppa:2022pnf} especially since the detection of gravitational waves (GW). A recent study \cite{Cheek:2022mmy} has looked into DM production from PBH mass and spin distributions where the PBHs are produced simultaneously at the moment when the PBH corresponding to the peak mass value is produced and also the evolution of the mass function is taken into account in \cite{Gehrman:2022imk}. In the scenario of a non-interacting DM production from the evaporation of a single monochromatic PBH distribution, a DM in the rather wide mass range of the order ($1-10^9$) GeV (a more specific range is mentioned in section \ref{sec:dmpheno}) cannot satisfy the relic ($\Omega h^2=0.120\pm0.001$ as per Planck data \cite{Planck:2018vyg}) in the PBH dominated region of parameter space \cite{Cheek:2021cfe} due to BBN constraints \cite{Sarkar:1995dd,Kawasaki:2000en,Hannestad:2004px,deSalas:2016ztq,DEBERNARDIS2008192}. An obvious question to ask in this context is  whether this rather strong constraint can be lifted in the presence of multiple PBHs with different initial masses. In this work,  we investigate the simplest possibility of DM being produced in the early universe from two separate monochromatic PBH mass distributions to check if at least a portion of this disallowed DM mass region can be redeemed. This would, of course, mean that a richer spectrum of PBHs can easily serve as the sole originator of the presently measured DM relic -- this observation serves as the motivation for the present work. We have done the analysis for a fermionic DM in this work and expect a qualitative extension of the results to the other types (i.e. scalar DM, vector DM etc.) of DM to be true.

The paper is organized as follows: in Sec.~\ref{sec:pbh}, we provide a quick review of PBH evaporation before discussing the DM phenomenology in Sec.~\ref{sec:dmpheno} and we offer our concluding remarks in Sec.~\ref{sec:conc}.


\section{Primordial Black Holes: A Review}
\label{sec:pbh}
PBHs have been of interest to physicists for decades. They can be created when the density fluctuations ($\frac{\delta \rho}{\rho}$) at the horizon level is greater than unity via what is commonly known as the Zel'dovich$-$Novikov mechanism \cite{Zeldovich:1967lct,Harrison:1969fb,Zeldovich:1972zz}. Other possible ways of formation are from topological defects \cite{Cotner:2016cvr} like the collapse of cosmic strings from second order phase transitions \cite{PhysRevD.43.1106, Kawana:2021tde, Vilenkin:2018zol}, inflationary perturbations \cite{Choudhury:2013woa}, or due to bubble collisions which arise from first order phase transitions in the early universe \cite{Baker:2021nyl, Hasegawa:2018yuy, Deng:2017uwc}. Interestingly, PBHs which originate from the collapse of highly over-dense regions have no upper bounds on their mass - such cases are generally studied by fitting into a mass spectrum \cite{Khlopov:2013ava}. This, however, is not true in the case when PBHs are formed from topological defects, because the mass of such PBHs is defined by the correlation length of the respective phase transition \cite{Rubin:2000dq,Martin:2019nuw}.

The study of PBHs is particularly important from a phenomenological point of view. PBHs with mass $M_{BH}^{in} \leq 10^9\rm{g}$ have evaporated well before the onset of the Big Bang Nucleosynthesis (BBN). In spite of this, their impact on the present day universe is quite strong. They can highly impact the baryon asymmetry of the universe \cite{Dolgov:2000ht}, the fraction of dark matter particles \cite{Chaudhuri:2020wjo,Chattopadhyay:2022fwa}, and would lead to the rise of density perturbations at relatively small scales \cite{Dolgov:2020xzo}.

The initial mass $M_{BH}^{in}$ of a PBH which is formed in the radiation dominated (RD) stage can be written in terms of the energy density of radiation $\rho_R$ \cite{Carr:2009jm,Carr:2020gox,Bernal:2020bjf}:
\begin{equation}
\label{eq:ini_mass}
M^{in}_{BH}=M_{BH}(T_{in})=\frac{4\pi}{3}\gamma \frac{\rho_{R}(T_{in})}{H^3(T_{in})},
\end{equation}
where $T_{in}$ is the temperature of the universe when the PBH was formed and $H(T_{in})$ is the value of the Hubble parameter at $T=T_{in}$. The numerical factor $\gamma$ depends on the gravitational collapse and is taken to be $\sim0.2$ for PBHs which are formed in the RD stage. The energy density of radiation - which is a function of temperature - takes the form
\begin{equation}
\rho_R(T)=\frac{\pi^2}{30}g_*(T)\,T^4,
\end{equation}
where $g_*(T)$ is the relative SM degrees of freedom  and is a function of temperature as well.  

As the PBH is formed, the temperature of the PBH is given by \cite{Hawking:1974rv,Hawking:1975vcx}:
\begin{equation}
	T_{BH}=\frac{M_p^2}{M_{BH}},
	\label{eqn:TvsM}
\end{equation}
where $M_p$ is the reduced Planck mass i.e. $M_p=\frac{1}{\sqrt{8\pi G}}$ with $G$ being the gravitational constant. Upon evaporation by Hawking radiation the PBH emits particles and the rate of mass loss of the PBH is governed by the following equation \cite{MacGibbon:1990zk,MacGibbon:1991tj,Perez-Gonzalez:2020vnz}:
\begin{equation} \label{decay}
	\frac{dM_{BH}}{dt}=-\sum_i\mathcal{E}_i(M_{BH})\frac{M_p^4}{M_{BH}^2}.
\end{equation}
Here, the evaporation function $\mathcal{E}_i$ depends on the mass of the $i^{th}$ particle as well as the mass of the PBH and Eqn.~\ref{decay} is summed over all the particle species. This early matter dominance in the form of PBH can distort the cosmological history. This in fact can modify the dark sector of the universe by injecting new dark matter particles by evaporation of the PBH. At the moment of PBH formation, the fraction of the PBH energy density to the total energy density denoted by $\beta$ is given by
\begin{equation} \label{beta}
	\beta=\frac{\rho_{BH}^{in}}{\rho_{BH}^{in}+\rho_{R}^{in}}\approx \frac{\rho_{BH}^{in}}{\rho_{R}^{in}}.
\end{equation}
The evolution history of the universe is also perturbed due to the modification of the Hubble parameter -- this is also mathematically represented by $\beta$ as shown in Eqn.~\ref{beta}
and an often-used rescaling of this parameter \cite{Cheek:2021cfe} is
\begin{equation}
\beta^\prime=\gamma^{\frac{1}{2}}\bigg(\frac{g_\star(T_{in})}{106.75}\bigg)^{-\frac{1}{4}}\beta\approx \gamma^{\frac{1}{2}}\beta.
\end{equation}
Dominance of either component is determined by comparing the value of $\beta$ with the critical value $\beta_c$, which is given by \cite{Masina:2020xhk}:
\begin{equation}
	\beta_c=\gamma^{-\frac{1}{2}}\left(\frac{\mathcal{G}g_{*,H}(T_{BH})}{10640 \pi}\right)^{\frac{1}{2}}\frac{M_{Pl}}{M^{in}_{BH}}.
\end{equation}
$\mathcal{G}$ is commonly known as the grey factor and $M_{Pl}$ is the Planck mass. If the value of $\beta$ is greater than $\beta_c$ the universe goes into the PBH (matter) domination at some stage in the early universe while $\beta<\beta_c$ implies the universe is radiation dominated throughout. The evaporation of the PBH is governed by the Boltzmann equation \cite{Bernal:2020bjf}:
\begin{equation}\label{eqn:rhobh}
	\frac{d\rho_{BH}}{dt}+3H\rho_{BH}=\frac{\rho_{BH}}{M_{BH}}\frac{dM_{BH}}{dt},
\end{equation}
where the rate of mass decay of the PBH $\frac{dM_{BH}}{dt}$ follows Eqn.~\ref{decay}.

An early matter domination might change the BBN predictions and hence, in order not to violate the well established results like the BBN temperature, the CMB results, and structure formation which took place at the later stage of the expansion of the universe, the PBH must evaporate before the onset of BBN and this necessitates the initial mass of the PBH to be bounded from above: $M^{in}_{BH}<10^9 \rm{g}$, as mentioned above. The lower bound on PBH mass arises from inflationary scales: $M_{BH}^{in}>10^{-1}\rm{g}$ \cite{Planck:2018jri}. Moreover, constraints from GW impose an upper bound on $\beta$ \cite{Domenech:2020ssp,Papanikolaou:2020qtd,Bernal:2020bjf}:
\begin{equation}
	\beta<\frac{10^9\rm{g}}{M_{BH}^{in}}10^{-4}.
\end{equation}

We now turn to understanding how some of the constraints might play out in the next-to-minimal scenario of two evaporating PBH distributions presenting a detailed phenomenological study of DM production in this case in Sec.~\ref{sec:dmpheno}. We propose a new way to relax the BBN constraint in the same scenario as well.


\section{Dark Matter Phenomenology}
\label{sec:dmpheno}
One of the key aspects of particulate dark matter that continues to elude us till date is its origin. In contrast to more traditional WIMP and FIMP scenarios, DM production from PBH evaporation does not require any interaction or portal between the SM sector and the DM candidate whereas the WIMP and FIMP mechanisms necessitate an interaction, however small. Thus the PBH evaporation as a DM production mechanism can explain the measured relic abundance for a DM particle interacting only gravitationally to the SM sector as has been assumed in this work. The key ingredients that constitute the parameter space of such DM production from a single non-spinning PBH distribution are $\beta^\prime$, $M_{BH}^{in}$, and the DM mass ($m_{DM}$). It is often convenient to represent the relic contours of DM in the $M_{BH}^{in}-\beta^\prime$ plane with each contour representing a different $m_{DM}$ value. While in the PBH dominated region ($\beta>\beta_c$), the relic is independent of $\beta^\prime$, in the radiation dominated region ($\beta<\beta_c$) the relic contours can have two different slopes depending upon the relative values of $m_{DM}$ and the temperature of the PBH ($T_{BH}$) \cite{Bernal:2020ili}. To radiate DM particles, the PBH temperature has to be higher than the DM mass. Hence, a PBH with small initial mass can easily produce DM particles due to its high $T_{BH}^{in}$. However, for a large $M_{BH}^{in}$, the corresponding $T_{BH}^{in}$ may not be large enough to compete with $m_{DM}$ (if $m_{DM}$ is relatively large) and hence cannot produce DM to begin with. But as the PBH evaporates through other SM states and loses mass, its temperature finally achieves a value high enough to start radiating DM particles as well.

\begin{figure}[h!]
\centering
\includegraphics[scale=0.5]{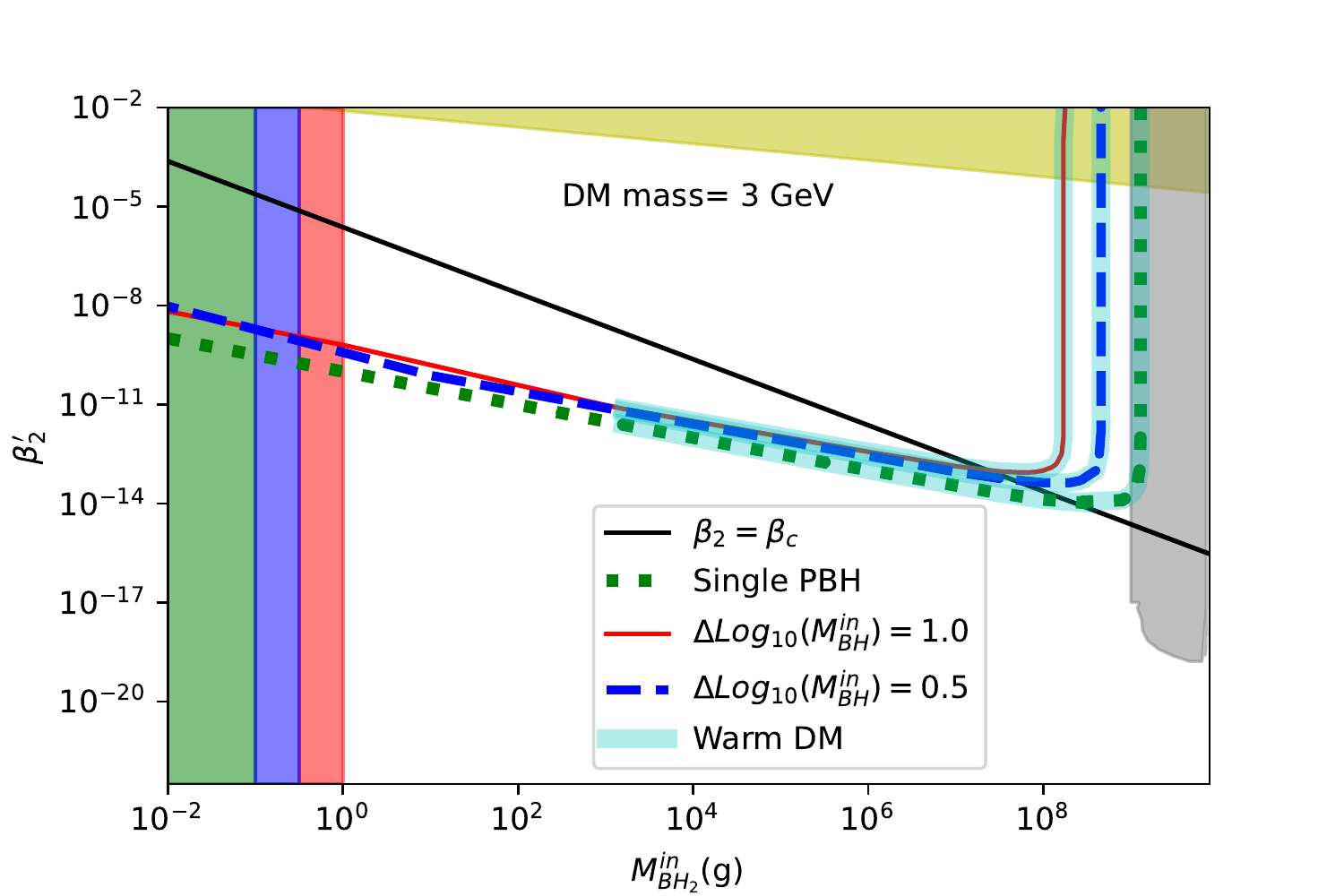}\includegraphics[scale=0.5]{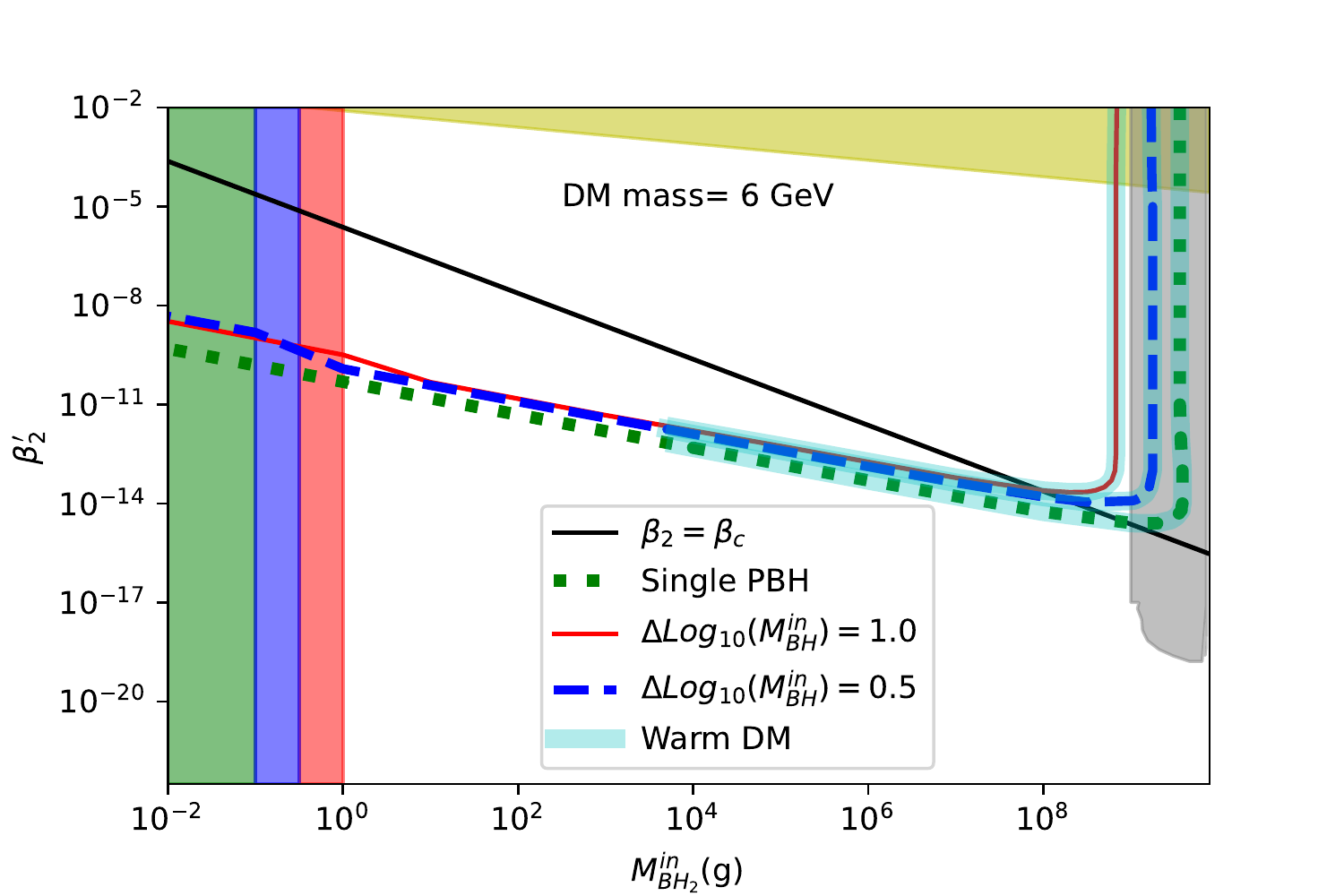}\ \includegraphics[scale=0.5]{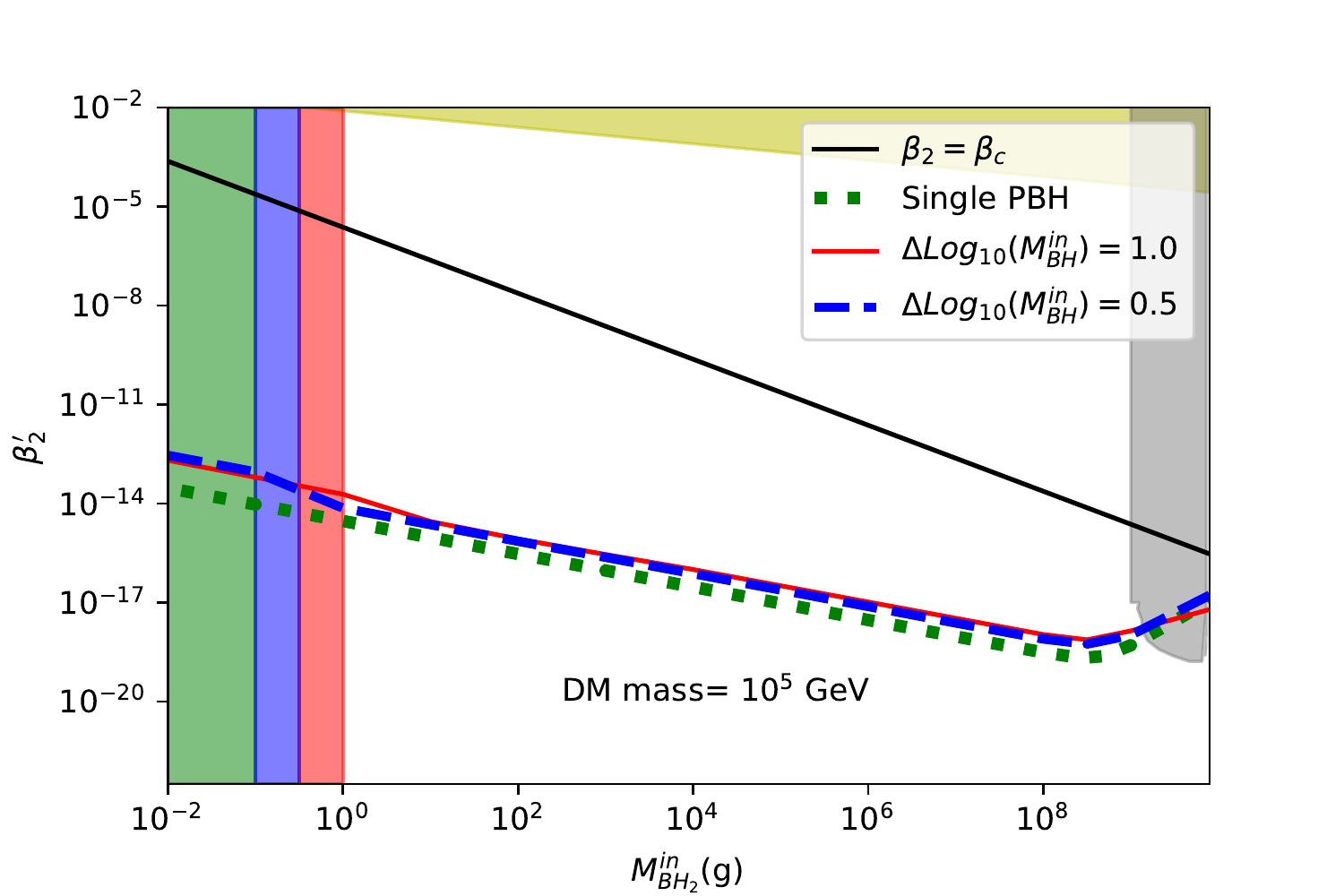}\includegraphics[scale=0.5]{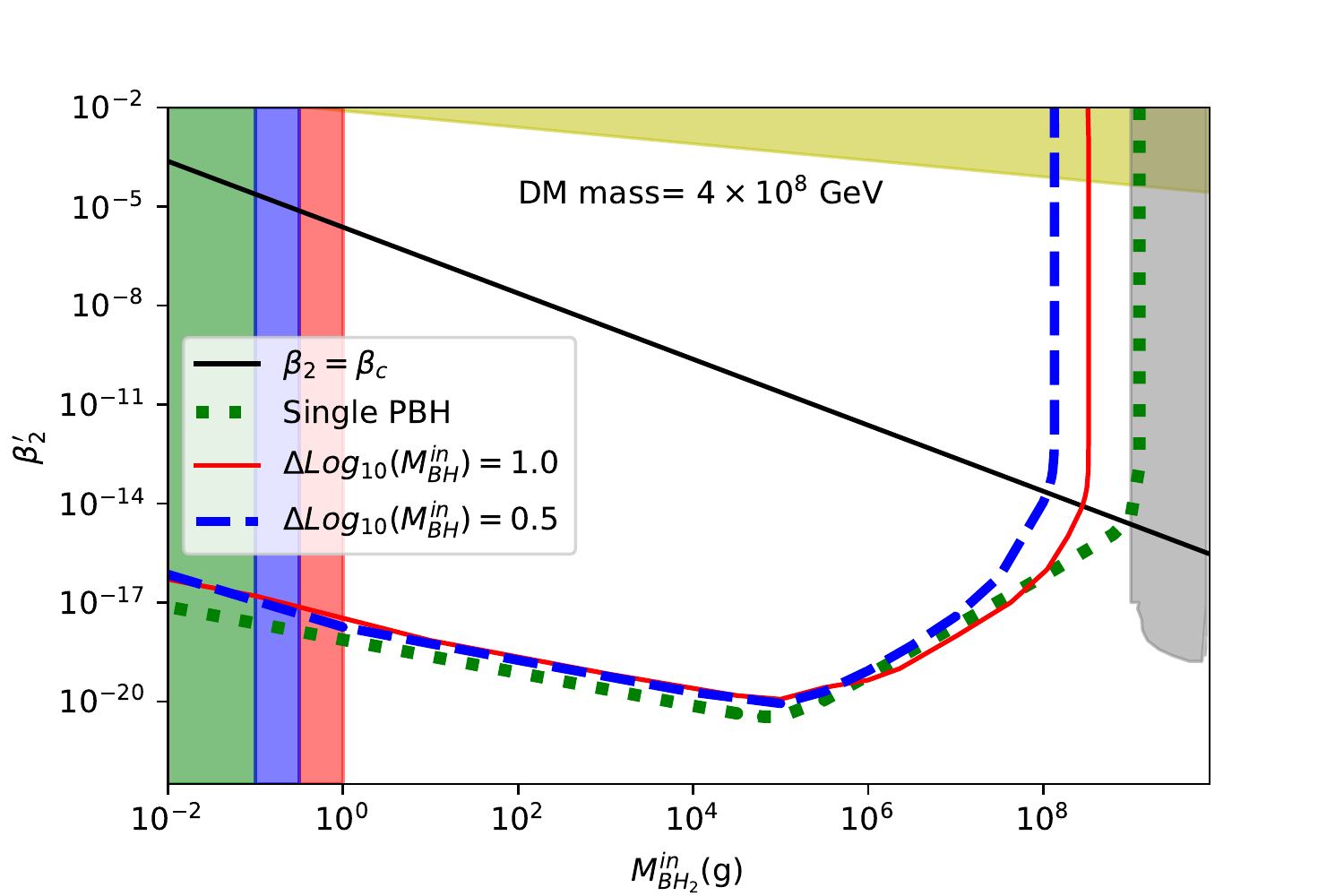}
\caption{Dark matter relic contours in the $M_{BH_2}^{in}(M_{BH}^{in})$-$\beta^\prime_2(\beta^\prime)$ parameter plane for DM masses 3 GeV (top left), 6 GeV (top right), $10^5$ GeV (bottom left) and $4\times10^8$ GeV (bottom right). The relic contours have been drawn for two choices of $Log_{10}(M_{BH_2}^{in})-Log_{10}(M_{BH_1}^{in})=$1.0 and  0.5 with $Log_{10}(\beta^\prime_2)-Log_{10}(\beta^\prime_1)=$ 1.0. The corresponding relic contours for the single-PBH scenario are also displayed. Exclusion bounds from BBN and GW valid for all cases are given by the light grey and light yellow shaded regions respectively. The inflation bound for a single-PBH scenario is given by the light green shaded region. For a two-PBH scenario the inflation bound becomes more severe as given by the additional light blue and light red shaded regions. The light cyan shade reflects the portion of the relic contour disallowed by the warm DM constraints.}
\label{fig:two_bh}
\end{figure}

Due to BBN constraints (discussed in Sec.~\ref{sec:pbh}) on the PBH distribution, the DM masses that are consistent with this scenario is limited as is evident in the literature \cite{Cheek:2021cfe}. For a fermionic DM in a PBH dominated scenario this forbidden DM mass region is given precisely by $(2.63-6.9\times10^8)$ GeV for a monochromatic mass distribution of PBH. It must also be noted that a DM particle of mass less than $\approx$ 1 TeV  is forbidden anyway by the warm DM constraints. A few example cases, spanned throughout the DM mass range in discussion, have been shown in Fig.~\ref{fig:two_bh} where in the PBH dominated region the relic contour for a single PBH distribution (given by the dotted green curve) falls inside the grey shaded region disallowed by BBN constraints\footnote{Of course, it is evident from Fig.~\ref{fig:two_bh} that this range is perfectly allowed in a \emph{radiation dominated} universe, but our goal here is to see if the same can be true in a \emph{matter dominated} universe as well. This is motivated by recent studies in non-standard cosmology wherein one can have an early matter dominated universe.}. The area above the contour is disallowed by DM overabundance. For a scenario with monochromatic PBHs, the GW and inflation bounds mentioned in Sec.~\ref{sec:pbh} are given respectively by the light yellow and light green shaded regions in Fig.~\ref{fig:two_bh}.  There is an interesting possibility of satisfying the relic in the PBH dominated region for a DM candidate of that forbidden mass range mentioned above with the existence of an additional PBH distribution with a different initial mass. The primary motivation for this is the idea that having more black holes in the spectrum at an early matter-dominated stage can affect the Hubble parameter, as their temporally separated evaporations enhance the DM energy density more than once and thus we should be able to account for the observed relic abundance of DM for lower initial PBH masses bringing the relic contour out of the BBN constrained region. However, the presence of two distributions comes with a few new degrees of freedom in the parameter space which demands to be defined properly and compared to the single monochromatic case.

\begin{figure}[h!]
\centering
\includegraphics[scale=0.6]{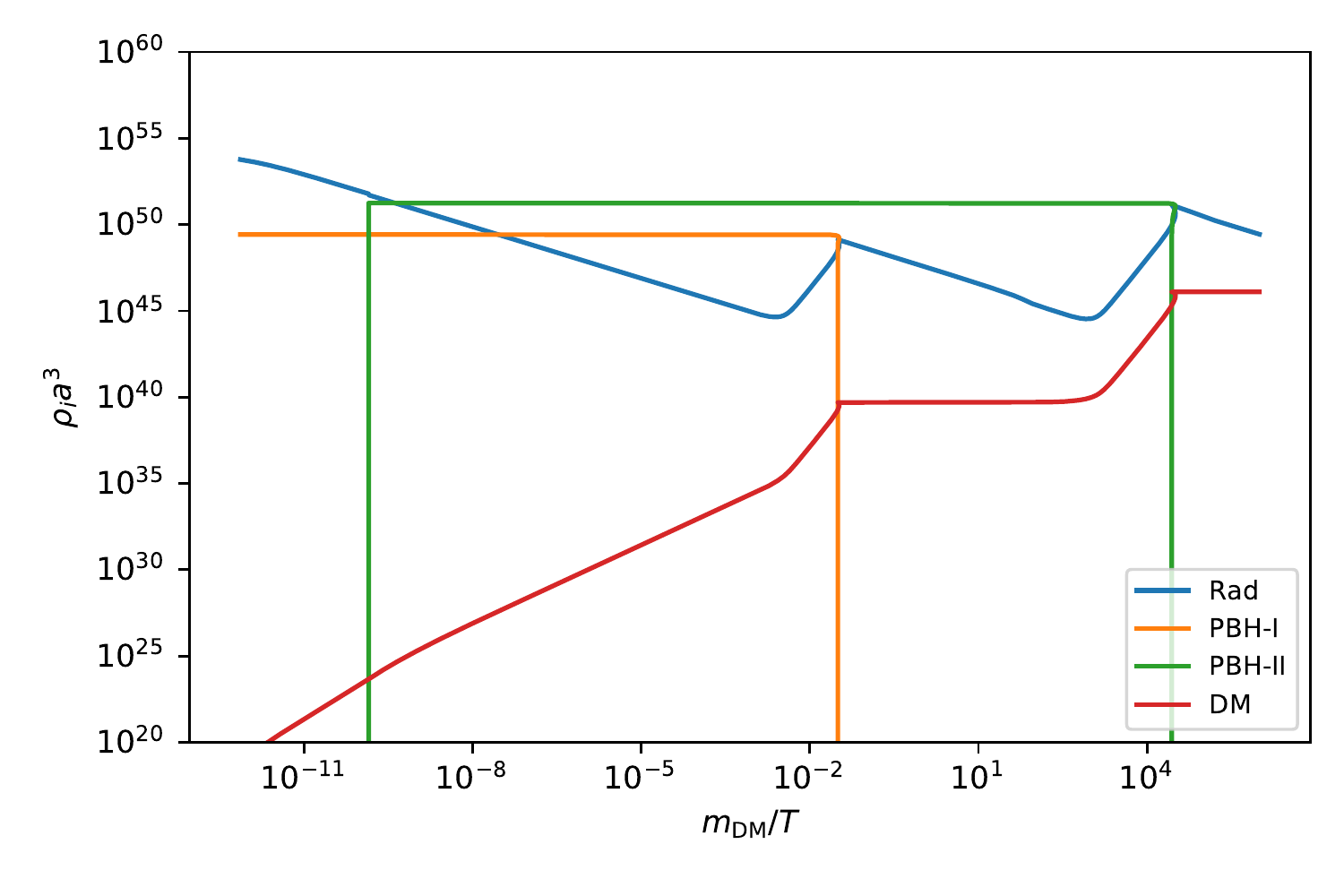}
\caption{Evolution of various components of the early universe for the following benchmark values: $M^{in}_{BH_1}=10^{5.0}$ g, $M^{in}_{BH_2}=10^{9.543}$ g, $\beta^\prime_1=10^{-5.0}$, $\beta^\prime_2=10^{-1.0}$ and $m_{DM}=10$ GeV. The relic abundance criteria is satisfied for these benchmark values. The evolution parameter is chosen to be inverse of the temperature of the universe ($T$) scaled appropriately by the DM mass.}
\label{fig:evol}
\end{figure}

Before taking forward this discussion further as a comparative study between the single-PBH and two-PBH scenarios, the details of a two-PBH scenario need to be formulated carefully. To start with, the timeline can be described in the following manner: in primordial times at temperature $T_1$, black holes of initial mass $M_{BH_1}^{in}$ are created and at a later time at temperature $T_2$, another set of black holes of initial mass $M_{BH_2}^{in}$ come into the picture. The obvious assumptions that go without saying are the conditions $T_1>T_2$  and hence, $M_{BH_1}^{in}<M_{BH_2}^{in}$ (see Eqn.~\ref{eq:ini_mass} with $\rho_{R}$ and $H$ replaced by their dependence on the temperature of the universe). Furthermore, now there will be two $\beta^\prime$ parameters that are defined as follows:
\begin{equation}
\beta_1^\prime=\sqrt{\gamma}\frac{\rho_{BH_1}(T_1)}{\rho_{BH_1}(T_1)+\rho_{R}(T_1)},\,\, \textrm{and}
\end{equation}
\begin{equation}
\label{equn:beta2p}
\beta_2^\prime=\sqrt{\gamma}\frac{\rho_{BH_1}(T_2)+\rho_{BH_2}(T_2)}{\rho_{BH_1}(T_2)+\rho_{BH_2}(T_2)+\rho_{R}(T_2)},
\end{equation}
where $\rho_{BH_i}$ denotes the energy density of the $i$-th PBH distribution and $\rho_{R}$ corresponds to the radiation energy density of the universe. Fig.~\ref{fig:evol} depicts one such scenario where there are two PBH distributions contributing to the radiation and DM relic as they evaporate via Hawking radiation. Here, we have plotted the evolution of co-moving energy densities of both the PBH distributions and DM along with the radiation component that is dimensionally appropriate to compare (co-moving energy density of radiation, $\rho_{R}a^4$, can not be compared with the co-moving matter). One can notice the co-moving energy density of DM increases as the PBHs evaporate with a sudden enhancement as the PBH-I comes close to complete evaporation. A similar trend is evident around the evaporation of the PBH-II. The radiation component also gets similarly elevated around those two regions. Once both the PBH distributions are evaporated completely the co-moving DM energy density stabilises as expected. We have used the ULYSSES code \cite{Cheek:2021cfe,Cheek:2021odj} to generate PBH contributions as input and then solved the Boltzmann equations mentioned in Appendix \ref{sec:equations} numerically to generate this plot.

The scenario with two monochromatic distributions of different initial mass involves more parameters as compared to the one with a single monochromatic PBH distribution. In order to draw a comparison between the two scenarios with mismatching number of parameters one will have to resort to a specific way of describing the scenario and the parameter space. Here, we have chosen to analyze the PBH-II of the two-PBH scenario in comparison to the PBH of the single-PBH scenario. The motivation behind this choice is to understand how the DM production from a PBH distribution (i.e. PBH-II) is modified in presence of another preexisting DM-producing PBH distribution (i.e. PBH-I). For that purpose we set the $\beta^\prime_2$ of the two-PBH scenario side by side with the $\beta^\prime$ of the single-PBH scenario and similarly contrast $M_{BH_2}^{in}$ against $M_{BH}^{in}$. In Fig.~\ref{fig:two_bh}, we have compared the relic contours of both single-PBH and two-PBH scenario for the same DM mass in the $M_{BH_2}^{in}(M_{BH}^{in})-\beta^\prime_2(\beta^\prime)$ plane. We have chosen four representative DM mass values from the range of interest: one ($4\times10^8$ GeV) near the upper end, two ($3$ GeV and $6$ GeV) in the lower end and the remaining one ($10^5$ GeV) around the middle of that range. We have fixed the excess degrees of freedom in the parameter space of the two PBH scenario by choosing $Log_{10}(\beta^\prime_2)-Log_{10}(\beta^\prime_1)=$ 1.0 and $Log_{10}(M_{BH_2}^{in})-Log_{10}(M_{BH_1}^{in})=$1.0, 0.5 as benchmark values. We find that for a fixed DM mass, having two PBH spectrums can significantly mitigate the BBN bound compared to the single-PBH scenario. It is evident in Fig.~\ref{fig:two_bh} that for the two-PBH case, the DM relic abundance criteria can be satisfied \emph{even in the PBH dominated region} in some cases which were disallowed in the single-PBH case because of the BBN bounds. For example, DM masses around the edges of the $1-10^9$ GeV region can now satisfy the relic with the addition of an extra PBH in the spectrum as has been demonstrated with DM masses of 3 GeV, 6 GeV and $4\times10^8$ GeV in Fig.~\ref{fig:two_bh}. However, for a similar logarithmic initial PBH mass difference and logarithmic $\beta^\prime$ difference, it can been seen that certain DM masses from the midrange are still forbidden by the BBN constraint even after the inclusion of an extra distribution of PBHs as has been shown in Fig.~\ref{fig:two_bh} for a DM mass of $10^5$ GeV. The precise DM mass range where DM relic can not be satisfied in the PBH dominated region now shrinks to $(7.24-8.91\times10^7)$ GeV for $\Delta Log_{10}(M_{BH}^{in})=1.0$ (in grams) and $(5.56-2.67\times10^7)$ GeV for $\Delta Log_{10}(M_{BH}^{in})=0.5$ (in grams). It has to be noted that the warm DM constraints \cite{Baldes:2020nuv,Baur:2017stq} become very much relevant for the lower DM mass range. The DM produced from the first monochromatic PBH mass distribution does not contribute sufficiently enough to the final DM relic to be constrained by the warm DM limit. However the energy density of the first distribution can still affect the evaporation temperature of the second one. Taking this fact into account, we have implemented the warm DM limits on the DM produced from the second PBH mass distribution and found that there is no significant change in the parameter space for bimodal production compared to the production from a single monochromatic PBH mass distribution. The whole PBH dominated region and a good portion of the radiation dominated region thus become disallowed for the DM masses of 3 GeV and 6 GeV regardless of whether produced from a single monochromatic distribution or from two such distributions as indicated by the light cyan shade in Fig.~\ref{fig:two_bh}.

Interestingly, the effect of the initial mass differences of PBH is not the same in different DM mass regions. In the higher DM masses, $m_{DM}$ is higher than $T_{BH_2}^{in}$ (the initial PBH-II temperature) and that leads to the opposite order of relic contours for $Log_{10}(M_{BH_2}^{in})-Log_{10}(M_{BH_1}^{in})$ values of 1.0 and 0.5 in the PBH dominated region compared to the low DM masses (where $m_{DM}<T_{BH_2}^{in}$ ) as can be seen by comparing the $m_{DM}=4\times10^8$ GeV graph of Fig.~\ref{fig:two_bh} with the $m_{DM}=3$ GeV graph. In the $m_{DM}=6$ GeV graph, the difference of  initial mass differences between the two monochromatic distributions plays a crucial role in determining whether or not the relic can be satisfied in the PBH dominated region. Thus we see a pattern emerging wherein a larger initial mass difference between the two black holes tends to accommodate smaller DM masses better while a smaller initial mass difference does a slightly better job of safeguarding more massive DM from the BBN bound. There are certain DM masses (as is illustrated in the $m_{DM}=10^5$ GeV plot) wherein neither of our two parameter choices help in overcoming the BBN bound. Another crucial aspect in this context is the addition of an extra component of PBH energy density both in the numerator and denominator in the definition of $\beta_2$ in the two-PBH scenario compared to the $\beta$ in the single-PBH case, which elevates the $\beta_2$ in comparison to the $\beta$ and thus reaching the PBH dominated region for comparatively lower initial mass of PBH. Even though we are mainly interested in the BBN constraints (light gray shaded region in the plots), the constraints from GW and inflation are also shown in the contour plots. It is imperative to remember that these constraints are applicable to both the PBHs. Since $M_{BH_1}^{in}$ is smaller than $M_{BH_2}^{in}$, PBH-I automatically satisfies the BBN bounds if satisfied by PBH-II. The same argument can be used for the GW bounds in the $M_{BH_1}^{in}$-$\beta^\prime_1$ plane which will already be satisfied once it is respected in the $M_{BH_2}^{in}$-$\beta^\prime_2$ plane\footnote{We have taken a simplified approach in applying the GW constraints on the two monochromatic PBH distributions separately in lieu of a detailed collective understanding of the two distributions.}. Hence, in Fig.~\ref{fig:two_bh} the GW and BBN bounds remain the same as the single-PBH scenario and are marked by light yellow and light grey shades respectively. However, PBH-II satisfying the inflation bound does not guarantee the same for PBH-I simply because $M_{BH_1}^{in}<M_{BH_2}^{in}$ as per our prescription. Hence, the inflation bound has to be applied on PBH-II in such a way that PBH-I automatically respects the bound as well. The three cases and corresponding contours are described as follows:
\begin{itemize}
\item For the single-PBH case the inflation bound demands that $M_{BH}^{in}>10^{-1}$ g. This bound is shown by the light green shade. To make it easier to comprehend, the relic contour corresponding to the single PBH scenario is shown by the same colour i.e. the green dotted curve.
\item For the two-PBH scenario with logarithmic initial mass difference of 0.5, the inflation bound needs to be modified because $M_{BH_2}^{in}=10^{-1}$ g already implies that $M_{BH_1}^{in}=10^{-1.5}$ g and thus $M_{BH_1}^{in}$ is in violation of the inflation bound. So, the inflation bound on $M_{BH_2}^{in}$ is modified in such a way that PBH-I also satisfies the bound. Hence, a region given by light blue shade is disallowed in addition to the green shaded region and the corresponding relic contour is given by the blue dashed curve.
\item Similarly for the initial logarithmic mass difference of 1.0, some more extra region will be disallowed which is given by the light red shaded part. In this case setting a lower bound on $M_{BH_2}^{in}$ at $10^0$ g makes sure that $M_{BH_1}^{in}$ respects the inflation bound of $10^{-1}$ g. Hence, the total region, consisting of the light green, the light blue and the light red shaded parts, is disallowed to account for both the PBHs in this case and the corresponding relic contour is given by the solid red curve.
\end{itemize}
It is to be noted that with the addition of an extra monochromatic PBH distribution in the spectrum, more parameter space opens up in the radiation dominated region compared to the single-PBH case.

\section{Conclusion}
\label{sec:conc}
PBH evaporation has furnished a method for the production of DM that is completely isolated from the SM sector. However, for an early PBH dominated era, the observed DM relic abundance can not be achieved for a rather large range of DM masses as it is forbidden by the BBN predictions. In this work, we have demonstrated that the presence of a second PBH distribution has the potential to mitigate this constraint - specifically,  employing a few benchmark values of DM mass from that forbidden region, we have shown that in the presence of another preexisting PBH distribution the relic abundance can be easily satisfied for DM masses around the edges of that forbidden region. However, the midrange DM masses still remain disallowed as has been shown for a specific benchmark value from midrange. This is a natural consequence of having one more PBH distribution that produces DM particles as well as an extra component of PBH energy density encapsulated in the definition of $\beta^\prime$ as $\beta^\prime_2$ in Eqn.~\ref{equn:beta2p}. Even though we have displayed the results by taking just a few DM masses as benchmark values, the results seem fairly interpolatable to a range of DM masses. Our analysis shows that some parts of the DM mass range, forbidden by BBN constraint from satisfying the relic in the PBH dominated region, can be redeemed with the help of this two-PBH scenario. However, the redemption of the whole region with just two monochromatic distributions seems difficult. A thorough study of a scenario with many PBHs with a fixed mass distribution spanned throughout the allowed PBH mass range produced at various time instances remains an interesting prospect in this context.


\appendix

\section{Relevant Evolution Equations}
\label{sec:equations}
The relevant Boltzmann equations along with Eqn.~\ref{eqn:rhobh} are the following \cite{Perez-Gonzalez:2020vnz}:
\begin{equation}
\frac{d\rho_R}{dt}+4H\rho_R=-\frac{\varepsilon_{SM}(M)}{\varepsilon(M)}\frac{1}{M}\frac{dM}{dt}\rho_{BH}, \,\textrm{and}
\end{equation}
\begin{equation}
\frac{dT}{dt}=-\frac{T}{\Delta}\bigg(H+\frac{\varepsilon_{SM}(M)}{\varepsilon(M)}\frac{1}{M}\frac{dM}{dt}\frac{g_\star(T)}{g_{\star s}(T)}\frac{\rho_{BH}}{4\rho_R}\bigg),
\end{equation}
where $\Delta=1+\frac{T}{3g_{\star s}(T)}\frac{dg_{\star s}(T)}{dT}$ and the evolution of the number density of DM from evaporating PBH is given by
\begin{equation}
\frac{dn_{DM}}{dt}+3Hn_{DM}=\frac{\rho_{BH}}{M_{BH}}\frac{dN_{DM}}{dt},
\end{equation}
 where $\frac{dN_i}{dt}$ is the emission rate of species $i$ from the evaporating PBH.

\begin{acknowledgements}
 BC acknowledges support from the Department of Science and Technology, India, under Grant CRG/2020/004171. AC thanks IIT-Gandhinagar for support. KL thanks Sujay Shil for helpful discussions.

\end{acknowledgements}

\bibliographystyle{apsrev}
\bibliography{Ref_Paper}
\end{document}